# Topological materials for full-vector elastic waves

Ying Wu[1*], Jiuyang Lu[1*], Xueqin Huang[1*], Yating Yang[1], Li Luo[1], Linyun Yang[2], Feng Li[3†], Weiyin Deng[1†], and Zhengyou Liu[4,5†]

[1]School of Physics and Optoelectronics and State Key Laboratory of Luminescent Materials and Devices, South China University of Technology, Guangzhou, 510640, China

[2]Department of Astronautic Science and Mechanics, Harbin Institute of Technology, Harbin, Heilongjiang 150001, China

[3]Centre for quantum physics, Key laboratory of advanced optoelectronic quantum architecture and measurement (MOE), School of Physics, Beijing Institute of Technology, Beijing, 100081, China

[4]Key Laboratory of Artificial Micro- and Nanostructures of Ministry of Education and School of Physics and Technology, Wuhan University, Wuhan 430072, China

[5]Institute for Advanced Studies, Wuhan University, Wuhan 430072, China

*Y.W., J.L. and X.H. contributed equally to this work
†Corresponding author. Email: phlifeng@bit.edu.cn; dengwy@scut.edu.cn; zyliu@whu.edu.cn

**Elastic wave manipulation is important in a wide variety of scales in applications including information processing in tiny elastic devices and noise control in big solid structures. The recent emergence of topological materials opens a new avenue toward modulating elastic waves in solids. However, because of the full-vector feature, and the complicated couplings of the longitudinal and transverse components of elastic waves, manipulating elastic waves is generally difficult, compared with manipulating acoustic waves (scalar waves) and electromagnetic waves (vectorial waves but transverse only). Up to date, topological materials, including insulators and semimetals, have been realized for acoustic and electromagnetic waves. Although topological materials of elastic waves have also been reported, the topological edge modes observed all lie on the domain wall. A natural question can be asked: whether there exists an elastic metamaterial with the topological edge modes on its own boundary only? Here, we report a 3D metal-printed bilayer metamaterial, insulating topologically the elastic waves. By**

**introducing the chiral interlayer couplings, the spin-orbit couplings for elastic waves are induced, which give rise to nontrivial topological properties. The helical edge states with the vortex feature are demonstrated on the boundary of the single topological phase. We further show a heterostructure of the metamaterial, which exhibits tunable edge transport. Our work may have potential in devices based on elastic waves in solids.**

Elastic metamaterials (EMMs), as one kind of artificially designed structures [1-3], are widely used to manipulate elastic waves in ways not found in nature, such as non-destructive testing [4], wave guiding [5], and information processing [6]. However, the wave transports in conventional EMMs suffer unavoidable backscattering in the presence of bends and defects. Recently, there are extensive efforts to explore the topological EMMs [7-8], which protect the high-efficiency transports of boundary modes, following the discovery of topological insulators in condensed-matter physics [9-12].

In two dimensions, there are two types of topological EMMs. The first hosts chiral edge states, in which the piezoelectric units break the time-reversal symmetry [13,14], mimicking the quantum anomalous Hall insulator. But the introduction of the active components [15] drastically increases the engineering complexity. Analogizing the quantum spin (valley) Hall insulator, the other possesses the helical boundary states with time-reversal symmetry, and has been realized in multi-scale EMMs [16-23]. Note that the full-vector feature of the elastic wave equation has yet to be fully considered in these systems. And the boundary states are usually localized at the domain walls between two distinct topological phases, acting as the interface states. It has remained an open question whether an EMM in the continuum can host the topological boundary states localized at the boundary of a single-phase, i.e., the edge states.

Here, we experimentally realize a topological EMM that inherently preserves helical edge states in a bilayer structure. Benefiting from the rich and flexible interlayer couplings, the bilayer structures provide new routes to explore the topological phase of matter [24-28]. Inspired by these works, we construct a bilayer structure, which has the chiral interlayer coupling to induce spin-orbit coupling for full-vector elastic waves and gives rise to nontrivial topological properties. Thanks to the advanced 3D metal-printing technology [29], the EMM with chiral interlayer coupling can be fabricated out with stainless steel in high reliability. This topological EMM provides a low-dissipation platform for elastic wave manipulation.

We start from a single layer EMM of a square lattice. As shown in Fig. 1a, a square block located at the center of the plate to modulate the elastic waves, giving rise to the band structure (Fig. 1b). Protected by the symmetry of $C_{4v}$ point group, a double degeneracy occurs determinately at the M point. The degeneracy is a typical quadratic Dirac point, featured by quadratic dispersions with opposite curvatures and $2\pi$ Berry phase around the point. The realization of topological insulator phases in the single EMM requires the quadratic Dirac degeneracy to be lifted by breaking parity or time-reversal symmetries [30]. For single-layer EMM, breaking parity symmetry hardly opens a direct band gap at the M point, while breaking time-reversal symmetry requires active components, which brings difficulty in fabrication.

We introduce interlayer couplings between two identical single-layer EMMs. Without couplings, the bilayer EMM has a copy of quadratic degeneracy at the M point. To open a band gap at that point, as depicted in Fig. 1c, the chiral interlayer couplings are introduced by four tilted pillars, which break parity symmetry. A thorough description around the M point can be obtained by an effective Hamiltonian deduced from a $k \cdot p$ perturbation theory. Considering all the crystalline symmetries, the linear part of the perturbation Hamiltonian vanishes, and meanwhile, the explicit form of the quadratic part is strictly constrained (see Supplementary Note 1), leading to the perturbation Hamiltonian as

$$\delta H = (\Delta k_x^2 + \Delta k_y^2)q_0 + 2\Delta k_x \Delta k_y q_1 \sigma_x + (\Delta k_x^2 - \Delta k_y^2)q_2 \sigma_z + \eta \tau_y \sigma_y. \qquad (1)$$

Here, Pauli matrices $\tau_i$ and $\sigma_i$ denote the layer pseudospin and the basis eigenmodes constituting the quadratic degeneracy in a single layer EMM, $(\Delta k_x, \Delta k_y)$ is the dimensionless wavevector deviating from the M point, and $q_i$ ($i = 0,1,2$) and $\eta$ represent the strengths of intralayer and interlayer couplings, respectively.

The perturbation Hamiltonian, consistent with the symmetries of the system, is capable of describing the band structures of the bilayer EMM. For $\eta = 0$, the perturbation Hamiltonian in Eq. (1) describes the single layer case. By fitting the dispersions around the M point of the single layer EMM, $q_i$ are determined as $q_0 =$ $0.38 \text{ Hz}^2$, $q_1 = 1.26 \text{ Hz}^2$, and $q_2 = 1.39 \text{ Hz}^2$. The fitting dispersions are plotted as black solid lines in Fig. 1b, showing consistent with the simulated ones (hollow circles). For $\eta \neq 0$, the last term in Eq. (1), representing the coupling between the layer pseudospins and the eigenmodes of a single layer, gives rise to the synthetic spin-orbit coupling for the bilayer EMM and produces a band gap at the M point (gray region in Fig. 1d). Determined by the width of the band gap, $\eta = 1.39 \text{ Hz}^2$. With $q_i$ and $\eta$, in the vicinity of the M point, the fitting curves (black solid lines in Fig. 1d) obtained from Eq. (1) capture well with the simulated dispersions for the bilayer EMM (hollow circles). Here the colors of the simulated dispersions (Figs. 1b and 1d) represent the weight of out-of-plane polarization with respect to total polarization (Methods). It shows that near the M point, the modes with different polarizations hybridize, polarizing partly in the out-of-plane direction. While in the frequency range of the band gap, the modes near the Γ point have hardly been disturbed and remain their in-plane features. This facilitates the experimental measurements by detecting the out-of-plane polarization in the frequencies of the band gap, where the distractions of modes far away from the M point are automatically blocked.

Whereas the elastic waves in the bilayer EMM are vectorial in nature, the EMM cannot be described by an equation that involves only the out-of-plane polarization but includes the in-plane ones. Such full-vector properties are essential for the effective Hamiltonian in Eq. (1), from which the topological properties of the bilayer EMM can be analytically deduced. After a similar transformation of a unitary matrix $U = -\frac{1}{2}\begin{pmatrix} 1 & -i \\ 1 & i \end{pmatrix} \otimes \begin{pmatrix} 1 & i \\ i & 1 \end{pmatrix}$, the Hamiltonian turns into a block-diagonal form $\delta H' = U\delta H U^{-1} = H_\uparrow \oplus H_\downarrow$, where the block matrices $H_{\uparrow/\downarrow} = (\Delta k_x^2 + \Delta k_y^2)q_0 + 2\Delta k_x \Delta k_y q_1 \sigma_x + (\Delta k_x^2 - \Delta k_y^2)q_2 \sigma_y \pm \eta \sigma_z$ representing the pseudospin up/down Hamiltonian. The two-block matrices share the identical bulk gap for nonzero $\eta$ and opposite Chern numbers $C_{\uparrow/\downarrow} = \pm \text{sgn}(\eta)$ for the first bulk band. A numerical approach to the topological properties can parallelly be performed by calculating the non-Abelian Wilson loop [31] based on the simulated data (see Supplementary Note 2), which again demonstrates the nontrivial topology of the bilayer EMM.

With the topologically nontrivial bulk properties, topological edge states are expected at the sample boundaries, regardless of the free and clamped ones. In contrast to the previously proposed topological boundary states for elastic waves, which propagates along the interface between two domains, the edge states here only require one single domain. We fabricate two samples with free and clamped boundaries, as shown respectively in Figs. 2a and 2d, each of which composes of $30 \times 7$ unit cells (see Methods and Supplementary Note 3). For the convenience and the efficiency, we excite the flexural motion by a piezoelectric actuator, which is glued at the free or the clamped boundaries of the sample (denoted by the green stars). Without doubt, the edge states can instead be excited by a source with in-plane vibration (see Supplementary Note 5). In the experiments, we detect the out-of-plane displacements fields ($w$) via a vertical laser vibrometer (Methods). The lower panels in Figs 2a and 2d show those along the two kinds of boundaries, edge states are excited and propagate along the boundaries compactly at a frequency of 26.75 kHz.

The experimental dispersions of the elastic edge states, isolated from the bulk, are presented in Figs. 2b and 2e. The color maps show the experimental results, and the solid curves represent the simulated ones. For the free boundary, the experimental and numerical results consistently show the existence of topological edge states in Fig. 2b. We note that a tiny band gap exists at $k_x = \pi/a$, due to the coupling of the pseudospin up and down states. The band gap is yet too small (0.3% in simulation) to be detected experimentally. For the clamped boundary condition, we obtain a pair of gapless edge states, denoted by the red and green curves in Fig. 2e, resulting in the pseudospin up and down edge states counter-propagating along the boundaries. Noted that only the edge state dispersions with positive group velocity are measured due to the sources located at the left sides. The simulated and measured results are in good agreement. Moreover, the gapless helical edge states can also exist at the 45° boundary, which is illustrated in Supplementary Note 4.

The behaviors of pseudospin-momentum locking of the helical edge states are reflected by the profiles of their eigenmodes both at the free and clamped boundaries. As shown in Figs. 2c and 2f, at different moments, the vortices of amplitudes give rise to the rotations of square blocks, whose directions lock to the edge state propagations, i.e., forward or backward. Specifically, the edge states propagating forward (backward) host the vortex in a clockwise (anticlockwise) direction. Therefore, the topological edge states are locked with the vortex directions. We further utilize multiple sources with different phases to selectively excite the pseudospin up and down edge states in Supplementary Note 5.

One distinct feature of topological edge states is their robust one-way transport along boundaries even with imperfection, such as sharp corners. To investigate this, we design a sample with a rectangular defect consisting of four $90°$ angles (Fig. 3a). The source is located at one end of the free boundary, which is composed of $34$ unit cells. We excite chirped signals with frequencies ranging from $23.5$ kHz to $30.5$ kHz. In Fig. 3b, we compare the measured transmission of the defect path with that of the straight sample of the same length. The slight difference between these two transmission curves within the topological gap (gray region), demonstrates weak backscattering of the edge state propagating along the rectangular defect. Figures 3c and 3d provide the field strengths $|w|$ at a frequency of $26.75$ kHz (dashed line in Fig. 3b). The experimental and simulated results show good agreement, confirming that the elastic wave propagates smoothly around the rectangular defect. Such robust transport of elastic waves along the clamped boundary is shown in Supplementary Note 5. Besides, it is worth pointing out that the 3D metal-printing samples [29] possess lower energy losses, advancing to design new functional devices for elastic waves.

Beyond a single phase of EMM, we further demonstrate that the transport path can be selected by combining two topologically distinct EMMs, paving a way to explore the devices, such as splitter and switch. View from the top, the unit cell with anti-clockwise (clockwise) chiral interlayer coupling is denoted as A (B). It can be predicted that a doubled number of edge states emerge at the interface between A and B, enabling the construction of complex networks for elastic wave manipulations (see Supplementary Note 6). As shown in Fig. 4a, we design a topological device including four ports formulated by A and B. The source is located at port 1 marked by a green star, and the right and left sides of the device are set as absorbing boundaries to avoid reflections. The width of the device is $W = 20a$, while the height $H$ can be manually tuned. We calculate the energy flow out of the ports 2-4, with the height $H$ ranging from $14a$ to $22a$, as shown in Fig. 4b. It is found that the transmissions of ports 3 and 4 show wane and wax with a period of $8a$ when the height changes. This oscillation feature comes from the interference of the two forward topological edge states belonging respectively to A and B (see Supplementary Note 6). Furthermore, because of spin-momentum locking, there is little energy flow to port 2.

Finally, we fabricate three representative samples of $H = 14a, 16a,$ and $18a$, to demonstrate the above path selection phenomenon in experiments. We observe that for $H = 14a$ almost all the elastic wave energy flows to port 3 (Fig. 4c). As the height increases to $H = 16a$, energy flows to both ports 3 and 4 (Fig. 4d). When $H = 18a$, nearly all the energy flows into port 4 (Fig. 4e). The experimental fields show good agreement with the simulations (Figs. 4f-4h), demonstrating that the transport of the elastic wave can be adjusted by the height of the device. Besides the free boundaries adopted here, samples with clamped boundaries are simulated in Supplementary Note 7 and show similar path selection properties.

In summary, we have proposed and realized a topological EMM possessing synthetic spin-orbit coupling that supports robust helical edge states. We experimentally demonstrate that the topological modes are available for reflection immune propagation, as well as flexible tunability of the elastic waves. The previous work [7] theoretically proposes an EMM hosting edge states, but it remains a huge challenge to observe the states in experiments due to the tiny bandgap. We anticipate our bilayer EMM to be a versatile platform to fulfill the applications for both the interface and surface waves, offering better flexibility for device realization. For example, the elastic edge states in EMM may find applications for splitters and switches in acoustic devices, enabling the construction of a monolithic elastic network. By stacking the structure layer by layer, our 2D topological EMM can be hopefully extended to 3D systems with intriguing topological transports, such as robust surface waves [32] and higher-order hinge states [33,34]. Furthermore, the proposed configuration can also be regarded as a bulk structure to investigate the topological defects, i.e., disclination [35, 36] or dislocation states [37, 38].

**Methods**

**Numerical simulations.** We performed the finite-element software to obtain all the simulations, using COMSOL Multiphysics with the Solid Structure module. We utilized an Eigenfrequency study with parameter sweep to calculate the band structures and the projected dispersions. We simulated the displacement fields with the low-reflection boundary conditions based on a Frequency Domain study. The largest and smallest element size was set to be $a/13$ and $a/20$, ensuring sufficient simulation accuracy. Each unit cell was meshed by 14586 elements and totally has 99159 degree-of-freedoms. For the material properties of the 316L stainless steel, we chose Young's modulus $E = 190$ Gpa, Poisson ratio $\mu = 0.26$, mass density $\rho = 8000$ kg/m$^3$.

**Sample preparation.** We fabricated the samples by using the 3D metal printing method, specifically the Selective Laser Melting technology. By leveraging the focusing laser spot, the prefabricated metal powder was fast melted, our samples made of 316L stainless steel with sophisticated shapes could be obtained directly, and additionally, the resulting density could reach more than 99%. After finishing printing, we polished the surface and removed the redundant supports, which were essential during the printing process (see Supplementary Note 2 for more details). The accuracy of the technology was up to 20 $\mu$m, which was suitable for our experimental samples.

**Experiments.** For the topological edge state measurements in Fig. 2, we used piezoelectric ceramic exciter as a point source to stimulate the elastic waves. Though the stimulation with vibration in out-of-plane direction, the excited waves contain both the in-plane and out-of-plane components. We applied ethylene-vinyl acetate copolymer to absorb extra elastic waves at the rest boundaries. We measured frequency responses of 30 scanned unit cells along the clamped and free boundaries, respectively. The amplitudes and phases were obtained by the Laser Vibrometer combined with a network analyzer. To obtain the dispersion curves of the boundary states, we measure the steady-state frequency response of the whole sample by spatially scanning with 1-mm steps (see Supplementary Note 2 for more details). To plot the dispersion curves, we performed a fast Fourier transformation of the $|w|\exp(i\varphi)$. To improve the resolution of the measure dispersions, the source was set at one side of the boundary, therefore, only the topological edge state with positive group velocity can be detected. Because of the time-reversal symmetry, the edge state with negative group velocity can be determined with mirror symmetry. We also applied the same experiment facilities by spatial scanning with 1mm-step along the *x*- and *y*-axis to characterize the fields of amplitude presented in Fig. 3 and 4.

For the transmission curves in Fig. 3b and 4b, we utilized a sweep frequency signal as a perturbation to excite the plate at the input location. The input (or output) signal can be obtained by a surface integral of the total energy flux over a rectangle cross profile containing four sites. Thus, the transmission can be given by $T(\omega) = 10\log(W_{\text{output}}/W_{\text{input}})$ or $T(\omega) = W_{\text{output}}/W_{\text{input}}$, in which W represent the total energy of elastic waves.

**Data availability**

The data that support the findings of this study are available from the corresponding author upon reasonable request.

**References**

1. Liu, Z., Zhang, X., Mao, Y., Zhu, Y. Y., Yang, Z., Chan, C. T. & Sheng, P. Locally resonant sonic materials. *Science* **289**, 1734-1736 (2000).

2. Frenzel, T., Kadic, M. & Wegener, M. Three-dimensional mechanical metamaterials with a twist. *Science* **358**, 1072-1074 (2017).

3. Ma, G. & Sheng, P. Acoustic metamaterials: from local resonances to broad horizons. *Sci. Adv.* **2**, e1501595 (2016).

4. Olsson III, R. H. & El-Kady, I. Microfabricated phononic crystal devices and applications. *Meas. Sci. Technol.* **20**, 012002 (2008).

5. Khelif, A., Choujaa, A., Benchabane, S., Djafari-Rouhani, B. & Laude, V. Guiding and bending of acoustic waves in highly confined phononic crystal waveguides. *Appl. Phys. Lett.* **84**, 4400-4402 (2004).

6. Sklan, S. R. Splash, pop, sizzle: Information processing with phononic computing. *AIP Adv.* **5**, 053302 (2015).

7. Matlack, K. H., Serra-Garcia, M., Palermo, A., Huber, S. D., & Daraio, C. Designing perturbative metamaterials from discrete models. *Nat. Mater.* **17**, 323-328. (2018).

8. Serra-Garcia, M., Peri, V., Süsstrunk, R., Bilal, O. R., Larsen, T., Villanueva, L. G., & Huber, S. D. Observation of a phononic quadrupole topological insulator. *Nature* **555**, 342-345. (2018).

9. Hasan M. Z. & Kane C. L. Colloquium: topological insulators. *Rev. Mod. Phys.* **82**, 3045-2067 (2010).

10. Qi, X. L. & Zhang, S. C. Topological insulators and superconductors. *Rev. Mod. Phys.* **83**, 1057-1110 (2011).

11. Kane, C. L. & Mele, E. J. Quantum spin Hall effect in graphene. *Phys. Rev. Lett.* **95**, 226801 (2005).

12. Bernevig, B. A., Hughes, T. L. & Zhang, S. C. Quantum spin Hall effect and topological phase transition in HgTe quantum wells. *Science* **314**, 1757-1761 (2006).

13. Darabi, A., Ni, X., Leamy, M., & Alù, A. Reconfigurable Floquet elastodynamic topological insulator based on synthetic angular momentum bias. *Sci. Adv.*, **6**, eaba8656 (2020).

14. Li, G. H., Ma, T. X., Wang, Y. Z., & Wang, Y. S. Active control on topological immunity of elastic wave metamaterials. *Sci. Rep.*, **10**, 1-8 (2020).

15. Darabi, A., Ni, X., Leamy, M. & Alù, A. Reconfigurable Floquet elastodynamic topological insulator based on synthetic angular momentum bias. *Sci. Adv.* **6**, eaba8656. (2020).

16. Cha, J., Kim, K. W. & Daraio, C. Experimental realization of on-chip topological nanoelectromechanical metamaterials. *Nature* **564**, 229-233 (2018).

17. Mousavi, S. H., Khanikaev, A. B. & Wang, Z. Topologically protected elastic waves in phononic metamaterials. *Nat. Commun.* **6**, 8682 (2015).

18. Miniaci, M., Pal, R. K., Morvan, B. & Ruzzene, M. Experimental observation of topologically protected helical edge modes in patterned elastic plates. *Phys. Rev. X* **8**, 031074 (2018).

19. Yu, S. Y., He, C., Wang, Z., Liu, F. K., Sun, X. C., Li, Z., Lu, H. Z., Lu, M. H., Liu, X. P. & Chen, Y. F. Elastic pseudospin transport for integratable topological phononic circuits. *Nat. Commun.* **9**, 613 (2018).

20. Chaunsali, R., Chen, C. W. & Yang, J. Subwavelength and directional control of flexural waves in zone-folding induced topological plates. *Phys. Rev. B* **97**, 054307 (2018).

21. Wu, Y., Chaunsali, R., Yasuda, H., Yu, K. & Yang, J. Dial-in topological metamaterials based on bistable Stewart platform. *Sci. Rep*. **8**, 112 (2018)

22. Yan, M., Lu, J., Li, F., Deng, W., Huang, X., Ma, J. & Liu, Z. On-chip valley topological materials for elastic wave manipulation. *Nat. Mater.* **17**, 993-998 (2018).

23. Vila, J., Pal, R. K. & Ruzzene, M. Observation of topological valley modes in an elastic hexagonal lattice. *Phys. Rev. B* **96**, 134307 (2017).

24. Lu, J., Qiu, C., Deng, W. Huang, X., Li, F., Zhang, F., Chen, S. & Liu, Z. Valley

topological phases in bilayer sonic crystals. *Phys. Rev. Lett.* **120**, 116802 (2018).

25. Deng, W., Huang, X., Lu, J., Peri, V., Li, F., Huber, S. D. & Liu, Z. Acoustic spin-Chern insulator induced by synthetic spin-orbit coupling with spin conservation breaking. *Nat. Commun.* **11**, 3227 (2020).

26. Peri, V., Song, Z. D., Serra-Garcia, M., Engeler, P., Queiroz, R., Huang, X., Deng, W., Liu, Z., Bernevig, A. B. & Huber, S. D. Experimental characterization of fragile topology in an acoustic metamaterial. *Science* **367**, 797-800 (2020).

27. Cao, Y., Fatemi, V., Fang, S., Watanabe, K., Taniguchi, T., Kaxiras, E. & Jarillo-Herrero, P. Unconventional superconductivity in magic-angle graphene superlattices. *Nature* **556**, 43-50 (2018).

28. Cao, Y., Fatemi, V. A., Fang, S., Tomarken, S. L., Luo, J. Y., Sanchez-Yamagishi, J. D., Watanabe, K., Taniguchi, T., Kaxiras, E., Ashoori R. C. & Jarillo-Herrero, P. Correlated insulator behaviour at half-filling in magic-angle graphene superlattices. *Nature* **556**, 80-84 (2018).

29. Zheng, X., Smith, W., Jackson, J. et al. Multiscale metallic metamaterials. *Nat. Mater.* 15, 1100–1106 (2016).

30. Chong, Y. D., Wen, X. G. & Soljačić, M. Effective theory of quadratic degeneracies. *Phys. Rev. B* **77**, 235125. (2008).

31. Yu, R., Qi, X. L., Bernevig, A., Fang, Z. & Dai, X. Equivalent expression of $Z_2$ topological invariant for band insulators using the non-Abelian Berry connection. *Phys. Rev. B* **84**, 075119 (2011).

32. Li, F., Huang, X., Lu, J., Ma, J. & Liu, Z. Weyl points and Fermi arcs in a chiral phononic crystal. *Nat. Phys.* **14**, 30-34 (2018).

33. Wei, Q., Zhang, X., Deng, W. Lu, J., Huang, X., Yan, M., Chen, G., Liu, Z. & Jia, S. Higher-order topological semimetal in acoustic crystals. *Nat. Mater.* **20**, 812-817 (2021).

34. Luo, L., Wang, H. X., Lin, Z. K., Jiang, B., Wu, Y., Li, F. & Jiang, J. H. Observation of a phononic higher-order Weyl semimetal. *Nat. Mater.* **20**, 794-799 (2021).

35. Liu Y., Leung S. W., Li F. F., Lin Z. K., Tao X. F., Poo Y. & Jiang, J. H. Bulk–

disclination correspondence in topological crystalline insulators. *Nature* **589**, 381–385 (2021).

36. Peterson, C. W., Li, T., Jiang, W., Hughes, T. L., & Bahl, G. Trapped fractional charges at bulk defects in topological insulators. *Nature*, **589**, 376-380 (2021).

37. Li, F. F. Wang H. X., Xiong Z., Lou Q., Chen P., Wu R. X., Poo Y., & Jiang J. H. Topological light-trapping on a dislocation. *Nat. Commun.* 9, 2462 (2018).

38. Roy, B., & Juričić, V. Dislocation as a bulk probe of higher-order topological insulators. *Phys. Rev. Research* **3**, 033107 (2021).

**Acknowledgements**

This work is supported by the National Natural Science Foundation of China (Nos. 11890701, 11974120, 11974005, 12074128), the Guangdong Basic and Applied Basic Research Foundation (Nos. 2019B151502012, 2021B1515020086, 2020A1515010549, 2021A1515010347), China Postdoctoral Science Foundation (NO. 2020M672615), and Guangzhou Basic and Applied Basic Research Foundation (No. 202102020349)

**Author contributions**

All authors contributed extensively to the work presented in this paper.

**Competing interests**

The authors declare no competing interests.

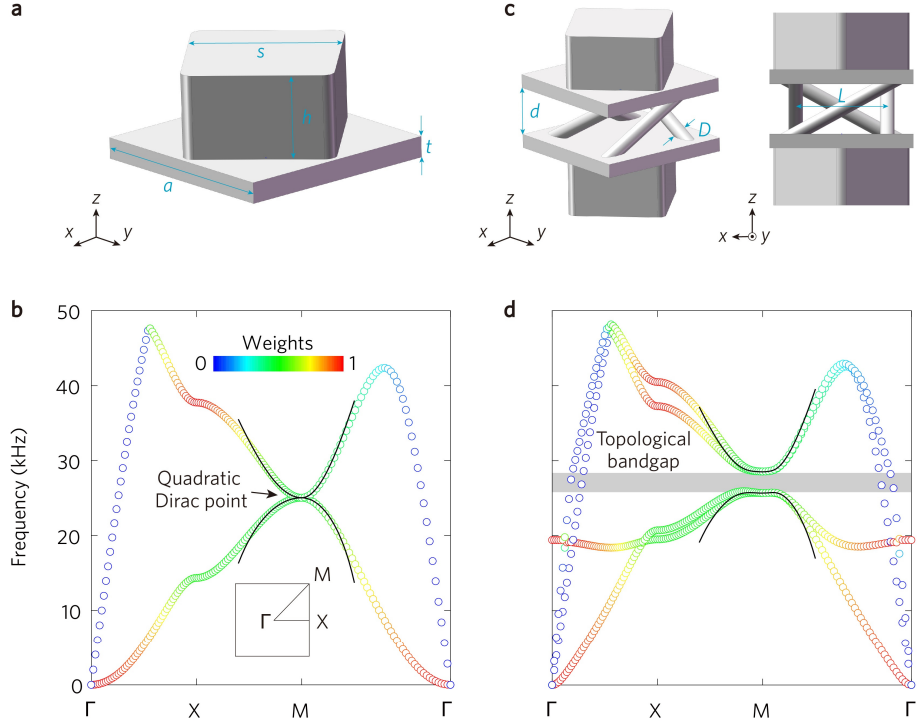

**Fig. 1 | Bulk dispersions of the single layer and bilayer EMMs. a**, Unit cell of the single layer EMM. **b**, Bulk dispersion of the single-layer. A quadratic degeneracy is preserved at the M point. The inset shows the first Brillouin zone. **c** and **d**, Unit cell and bulk dispersion of the bilayer EMM. Chiral interlayer coupling is introduced by four tilted pillars, which opens a topologically nontrivial band gap (the gray region in **d**). In **b** and **d**, the color maps represent the proportion of the out-of-plane displacement $|w|$, and the black solid curves denote the fitting data of the effective Hamiltonians. The geometry parameters are: lattice constant $a = 10.0$ mm, plate thickness $t = 1.0$ mm, length (height) of square block $s = 7.0$ mm ($h = 4.0$ mm), distance between two layers $d = 3.5$ mm, the diameter of the tilted pillar $D = 1.0$ mm, and distance between two opposite titled pillars $L = 6.4$ mm.

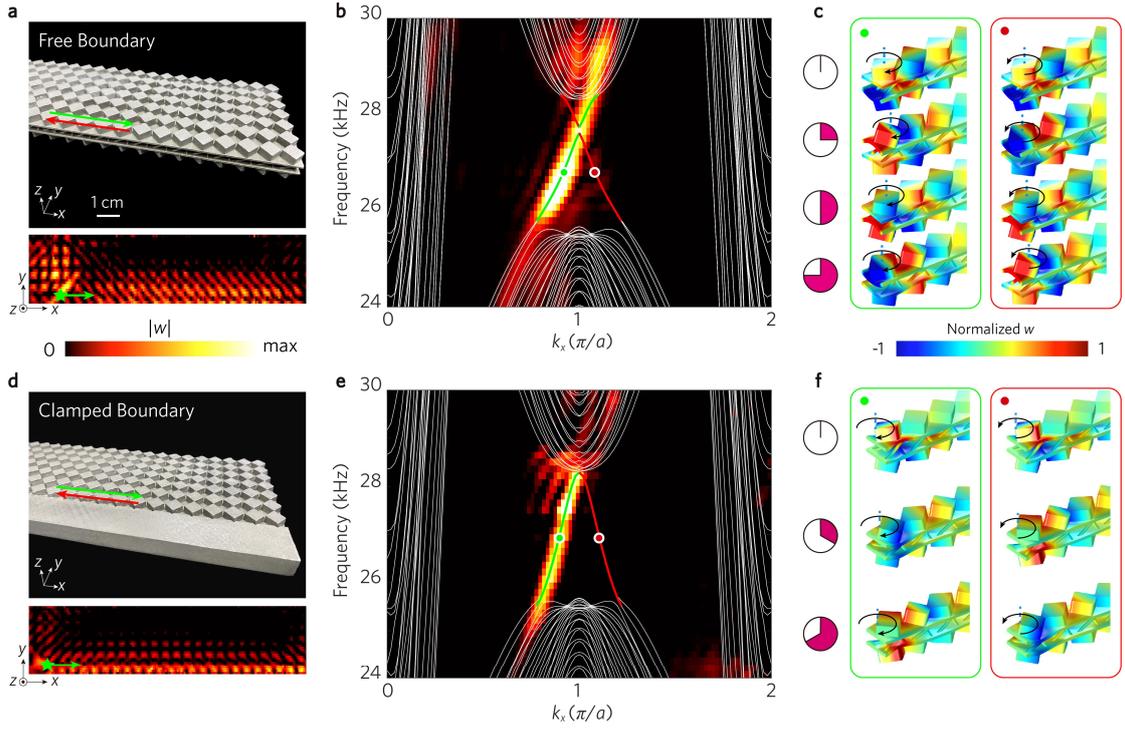

**Fig. 2 | Measured helical edge states on the free and clamped boundaries. a**, Upper panel: photograph of the 3D metal-printing sample for the free boundary. Lower panel: Measured distribution of $|w|$ along the free boundary. Green star: a point source. **b**, Measured (color map) and simulated (curves) dispersions projected on the free boundary. The red and green curves denote dispersions of the edge states counter-propagating along the free boundary (illustrated by red and green arrows in **a**). **c**, Simulated field distributions of the edge states, marked by the circles in **b**, at sequential instants. The black arrows indicate the rotation of the square blocks around their centers (blue dashed lines). **d-f**, the same as in **a-c**, but for the clamped boundary.

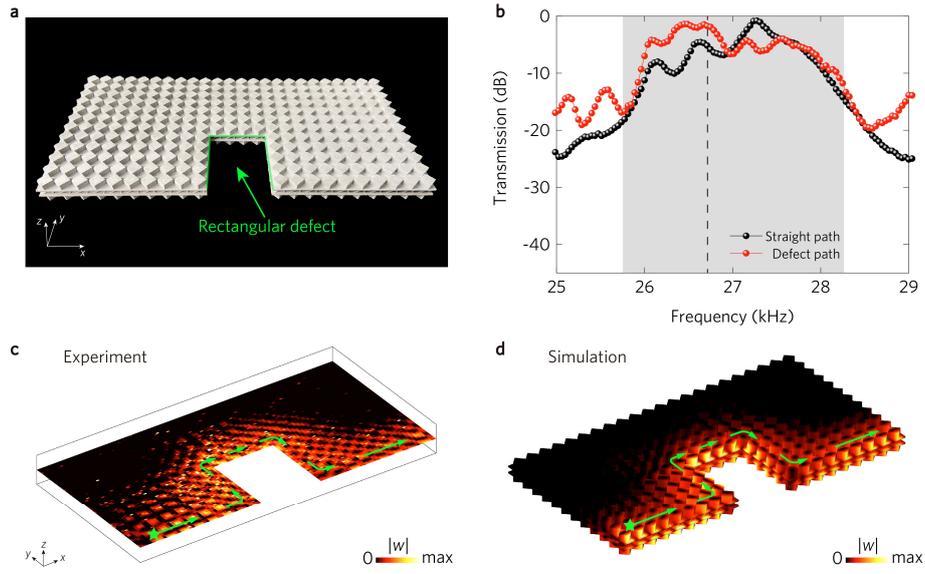

**Fig. 3 | Robust transport of the elastic edge states in the presence of a rectangular defect. a**, Photograph of the EMM sample with a rectangular defect along the free boundary. **b**, Measured transmission for the defect path (red dots), compared with that for a straight path (black dots). The gray region denotes the bulk band gap. **c** and **d**, Measured and simulated distribution of |$w$| at 26.75 kHz, respectively. Green stars denote the sources.

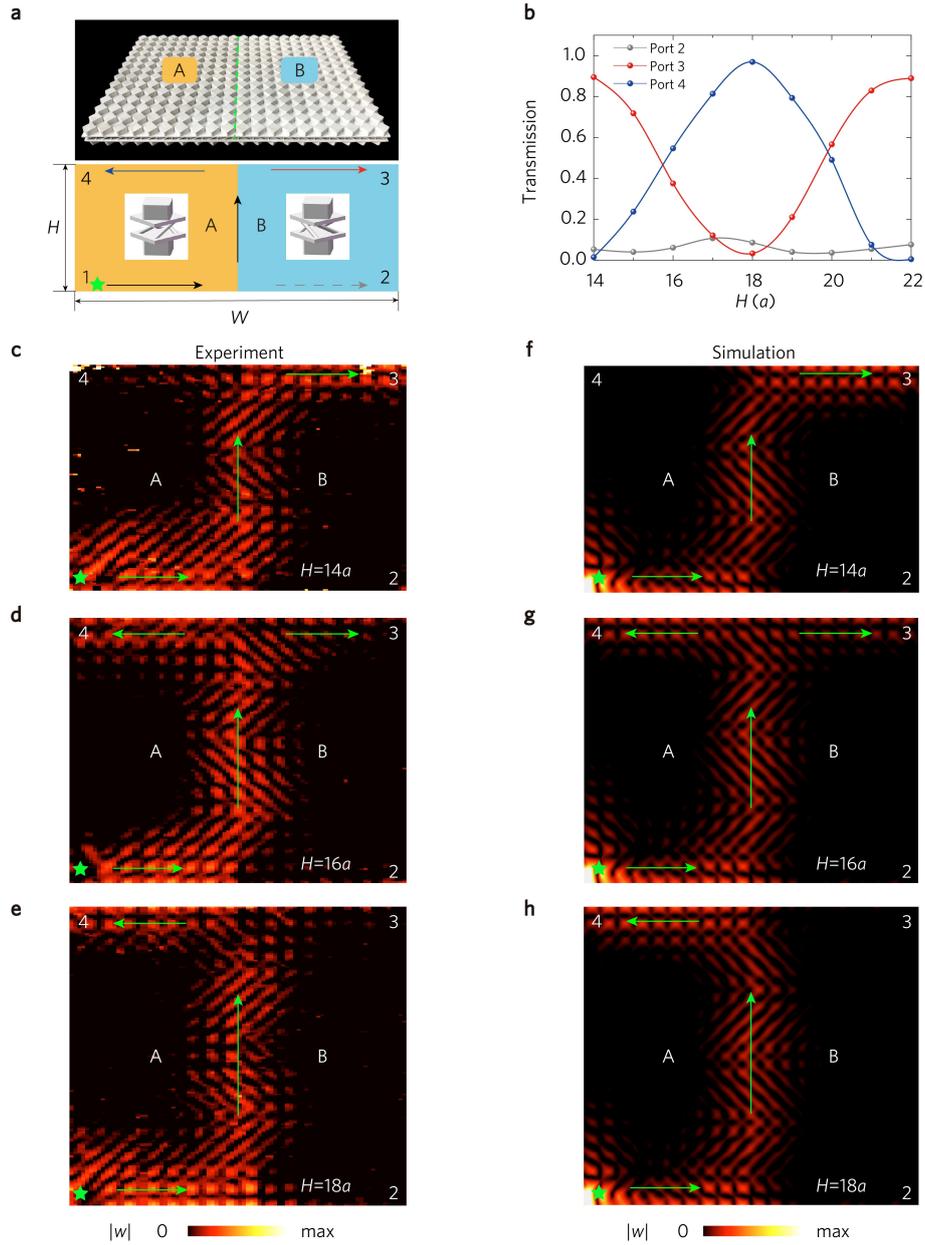

**Fig. 4 | Selective transport at topological channels for elastic edge states. a**, Photograph of the sample, composed of A and B structures. There exist four ports in the device including one input, denoted as 1, and three output ports, denoted as 2, 3, and 4, respectively. The height and width of the sample are denoted as $H$ and $W$. **b**, Transmissions in output ports 2-4 with respect to $H$ at 26.75 kHz. **c-h,** Measured and simulated $|w|$ for $H = 14a$, $16a$, and $18a$, respectively. The green stars denote the point sources. The green arrows show the directions of edge state propagation.